\documentclass[conference]{IEEEtran}
\usepackage{amsmath,graphicx,amsfonts,epsfig}
\usepackage{times}
\usepackage{amsbsy}
\usepackage{latexsym}
\usepackage{amssymb}
\usepackage{color}

\begin{document}

\title{Dynamic Topology Adaptation and Distributed Estimation for Smart Grids}

\author{\IEEEauthorblockN{Songcen Xu}
\IEEEauthorblockA{Communications Research Group\\Department of Electronics\\
University of York, U.K.\\
Email: sx520@york.ac.uk}
\and
\IEEEauthorblockN{Rodrigo C. de Lamare}
\IEEEauthorblockA{Communications Research Group\\Department of Electronics\\
University of York, U.K.\\
Email: rcdl500@york.ac.uk}
\and
\IEEEauthorblockN{H. Vincent Poor}
\IEEEauthorblockA{Dept. of Electrical Engineering\\Princeton University\\
Princeton, NJ 08544, USA\\
Email: poor@princeton.edu}}

\maketitle

\begin{abstract}
This paper presents new dynamic topology adaptation strategies for
distributed estimation in smart grids systems. We propose a dynamic
exhaustive search--based topology adaptation algorithm and a dynamic
sparsity--inspired topology adaptation algorithm, which can exploit
the topology of smart grids with poor--quality links and obtain
performance gains. We incorporate an optimized combining rule, named
Hastings rule into our proposed dynamic topology adaptation
algorithms. Compared with the existing works in the literature on
distributed estimation, the proposed algorithms have a better
convergence rate and significantly improve the system performance.
The performance of the proposed algorithms is compared with that of
existing algorithms in the IEEE 14--bus system.
\end{abstract}


\section{Introduction}
The electric power industry is likely to involve many more fast
information gathering and processing devices (e.g., phasor
measurement units) in the future, enabled by advanced control,
communication, and computation technologies \cite{Bose}. As a
result, the need for more decentralized estimation and control in
smart grids systems will experience a high priority. Several works
in the literature have proposed strategies for distributed
estimation \cite{Korres,Xie1,Xie}. With existing algorithms, the
neighbors for each bus are fixed. When there are links that are more
severely affected by noise or other disturbances, these approaches
may not provide an optimized estimation performance for each
specified bus. Moreover, with the number of neighbor buses
increasing, each bus requires a large network bandwidth and transmit
power. Therefore, a key problem with the strategies reported so far
in the literature is that they do not exploit the topology of the
smart grids system and the knowledge about the poor links to improve
the performance of distributed estimation techniques.

The objective of this paper is to propose fully distributed dynamic
topology adaptation algorithms for distributed estimation in smart
grids system, in order to optimize the performance and minimize the
mean-square error (MSE) associated with the estimates. We propose
two dynamic topology adaptation strategies, the proposed algorithms
exploit the knowledge about the poor links and the topology of the
system to select a subset of links that results in an improved
estimation performance. For the first approach, we consider a
dynamic exhaustive search--based topology adaptation (DESTA)
strategy. For the DESTA algorithm, we consider all possible
combinations for each bus with its neighbors. Then we choose the
combination associated with the smallest MSE value.

In the second approach, we introduce the dynamic sparsity--inspired
topology adaptation (DSITA) algorithm. A reweighted zero attraction
(RZA) strategy is incorporated into the dynamic topology adaptation
algorithm. The RZA approach is usually employed in applications
dealing with sparse systems in such a way that it shrinks the small
values in the parameter vector to zero, which results in better
convergence rate and steady--state performance. Different from prior
work with sparsity--aware algorithms
\cite{Chen,Rcdl1,Xu1,zhaocheng12}, the proposed DSITA algorithm
exploits the possible sparsity of the MSE associated with each of
the links in a different way and employs the Hastings rule
\cite{Hastings}. The DSITA shrinks to zero the links that have a
poor performance. To implement DSITA, we introduce a convex penalty,
i.e., an $\ell_1$--norm term to adjust the combination coefficients
for each bus with its neighbors, in order to select the neighbor
buses that yield the smallest MSE values.

The dynamic topology adaptation is achieved as follows:
\begin{itemize}
\item For a specified bus, we calculate the MSE at all its neighbor
buses including the specified bus itself through the previous estimate.
\item For the bus with the maximum MSE, we impose a penalty and give a reward to the bus with the smallest MSE.
\end{itemize}
The proposed DSITA algorithm performs this process automatically. By
using the DSITA algorithm, some buses with unsatisfactory
performance will be eliminated and some poor buses will be taken
into account when their performance improves, which means the system
topology will change automatically as well. To further improve the
performance of distributed estimation techniques, we consider the
Hastings rule \cite{Hastings} to construct the initial combination
coefficients and incorporate it into the proposed algorithms.

This paper is organized as follows. Section II describes the system
model and the problem statement. In section III, the proposed
dynamic topology adaptation algorithms are introduced. The numerical
simulation results are provide in section IV. Finally, we conclude
the paper in section V.

Notation: We use boldface uppercase letters to denote matrices and
boldface lowercase letters to denote vectors. We use $(\cdot)^{-1}$
to denote the inverse operator, and $(\cdot)^*$ for conjugate
transposition.

\section{System Model and Problem Statement}

\begin{figure}[!htb]
\begin{center}
\def\epsfsize#1#2{0.825\columnwidth}
\epsfbox{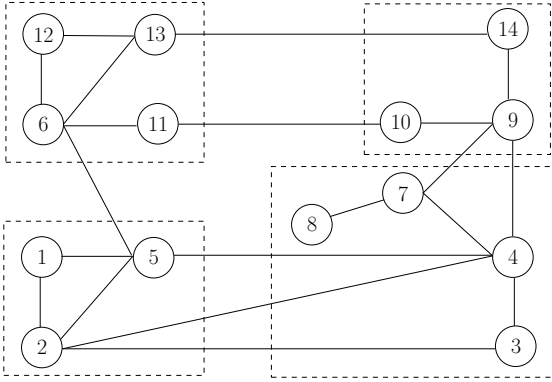} \caption{\footnotesize
IEEE 14--bus system}
\vskip -5pt
\label{fig1}
\end{center}
\end{figure}

We consider an IEEE 14--bus system \cite{Bose}, where 14 is the
number of substations. At every time instant $i$, each bus $k,k=1,2,
\ldots, 14 ,$ takes a scalar measurement $z_k(i)$ according to
\begin{equation}
{z_k(i)}= {H_k (\boldsymbol x(i))+ e_k(i)},~~~
k=1,2, \ldots, 14 \label{z_k},
\end{equation}
where $\boldsymbol x(i)$ is the state vector of the entire
interconnected system, $H_k({\boldsymbol x(i)})$ is a nonlinear
measurement function for bus $k$. The quantity ${e_k(i)}$ is the
measurement error with mean equal to zero and which corresponds to
bus $k$. Fig. \ref{fig1} shows a standard IEEE--14 bus system with
four nonoverlapping control areas.

Initially, we focus on the linearized DC state estimation problem.
The system is built with 1.0 per unit (p.u) voltage magnitudes at
all buses and j1.0 p.u. branch impedance. Then, the state vector
$\boldsymbol x(i)$ is taken as the voltage phase angle vector
$\boldsymbol \theta$ for all buses. Therefore, the nonlinear
measurement model for state estimation (\ref{z_k}) is modified to
\begin{equation}
{z_k(i)}= {\boldsymbol h_k(i)^*\boldsymbol \theta+ e_k(i)},~~~
k=1,2, \ldots, 14 .
\end{equation}
where $\boldsymbol h_k(i)$ is the measurement Jacobian vector for
bus $k$. Then, the aim for the distributed estimation algorithm is
to compute an estimate of $\boldsymbol \theta$, which can minimize
the cost function
\begin{equation}
{J_{x}({\boldsymbol x})} = {\mathbb{E} |{ z_k(i)}- {\boldsymbol h_k(i)}^*{\boldsymbol x_k(i)}|^2} \label{cost function},
\end{equation}
where $\mathbb{E}$ denotes the expectation operator.

A LS--type distributed algorithm, named Modified--Coordinated State
Estimation ($\mathcal{M}$--$\mathcal{CSE}$), has been reported in
the literature \cite{Xie}. In this strategy, the system is
decomposed into N areas. Based on the current state vector
$\boldsymbol x_n(i)$, where n=1,2, \ldots, N, the exchanged data
$\{\boldsymbol x_l(i) \}_{l\in\Omega_n}$, and the measurement vector
$\boldsymbol  z_n$, the estimate of the state at the $n$th control
area can be updated via the following formula
 \begin{equation}
\begin{split}
{\boldsymbol x_n{(i+1)}}&= {\boldsymbol x_n{(i)}}-[\beta(i)\sum\limits_{l\in \Omega_n}(\boldsymbol x_n{(i)}-\boldsymbol x_l{(i)}) \\
&\ \ \ \ \ \ \ \ \ \ \ \ \ -\alpha(i) \boldsymbol H_n^T(\boldsymbol z_n-\boldsymbol H_n\boldsymbol x_n(i))],
\end{split}
\end{equation}
where $\{\alpha(i)\}$, $\{\beta(i)\}$ are time--varying weight sequences.

For the existing strategies in the literature for smart grids, the
system communication topology is fixed. This situation will cause a
problem when some of the neighbor buses have a poor performance, or
the links between buses experience a disturbance. Also, there is no
chance for the bus to discard the poorly performing neighbors rather
than continue to use their information. In order to solve these
problems and optimize the distributed estimation process, we need to
provide the system with the ability to adapt the topology
dynamically.

\section{Proposed Dynamic Topology Adaptation Strategies}

In this section, we introduce dynamic topology adaptation strategies
for distributed estimation in smart grids. The aim of our proposed
DESTA and DSITA algorithms is to optimize the distributed estimation
process and improve the performance of the smart grids. These two
algorithmic strategies give the buses the ability to choose their
neighbors based on their MSE performance. Note that other
performance criteria are possible.

\subsection{Hastings Rule}

We first describe a combination rule -- Hastings rule that has an
improved performance as compared to the Metropolis rule
\cite{Hastings}, and is incorporated into the proposed algorithms.
The combination coefficient $c_{kl}$ for a bus $k$ and its neighbor
bus $l$, can be calculated under the Hastings rule as follows
\begin{equation}
c_{kl}=\left\{\begin{array}{ll}
\frac{\sigma_{n,k}^2}
{max\{|\mathcal{N}_k|\sigma_{n,k}^2,|\mathcal{N}_l|\sigma_{n,l}^2\}},\ \ $if\  $k\neq l$\ are linked$\\
1 - \sum\limits_{l\in \mathcal{N}_k / k} c_{kl}, \ \ \ \ \ \ \ \ \ \ \  $for\  $k$\ =\ $l$$
\end{array}
\right. \label{hast_rule}
\end{equation}
where $|\mathcal{N}_k|$ denotes the cardinality of $\mathcal{N}_k$, and $\sigma_{n,k}^2$ stands for the noise variance on bus $k$. All $c_{kl}$ should satisfy
\begin{equation}
\sum\limits_{l} c_{kl} =1 , l\in \mathcal{N}_k \forall k.
\label{ctotal 1}.
\end{equation}
The Hastings rule is a fully--distributed solution, as each bus $k$ only needs to obtain the degree--variance product $(|\mathcal{N}_l|-1)\sigma_{n,l}^2$ from its neighbour $l$, to get the combination coefficient \cite{Zhao}.

\subsection{Dynamic Exhaustive Search--Based Topology Adaptation (DESTA)}

In the proposed DESTA algorithm, we divide the distributed
estimation process into two steps. The first step is the adaptation
step and the second step is the combination step. For the proposed
DESTA algorithm, we employ the adaptation strategy given by
\begin{equation}
{\boldsymbol {\psi}}_k(i)= \boldsymbol x_k(i-1)+{\mu}_k {\boldsymbol
h_k(i)}[{ z_k(i)}- {\boldsymbol h_k(i)^*}{\boldsymbol x_k(i-1)}]^*.
\label{adap step}
\end{equation}
Following the adaptation step, we introduce the combination step for
the DESTA algorithm, based on an exhaustive search strategy. At
first, we introduce a tentative set $\Omega_s$ using a combinatorial
approach described by
\begin{equation}
{\Omega_s}\triangleq {C_{T}^t},~~~t=1,2, \ldots, T,
\end{equation}
where $\{T\}$ is the total number of buses linked to bus $k$
including bus $k$ itself. This combinatorial strategy will cover all
combination choices for each bus $k$ with its neighbors. After the
tentative set $\Omega_s$ is defined, we redefine the cost function
(\ref{cost function}) for each bus as
\begin{equation}
{J_{\psi}({\boldsymbol \psi})} \triangleq {\mathbb{E} |{ z_k(i)}-{\boldsymbol h_k(i)}^*{\boldsymbol \psi}|^2} ,
\end{equation}
where
\begin{equation}
{\boldsymbol \psi} \triangleq \sum\limits_{l\in \Omega_s} c_{kl} \boldsymbol\psi_l(i)
\end{equation}
Then, we introduce the error pattern for each bus, which is defined
as
\begin{equation}
{e_{\Omega_s}(i)} \triangleq { z_k(i)}-{\boldsymbol h_k(i)}^*{[\sum\limits_{l\in \Omega_s} c_{kl} \boldsymbol\psi_l(i)]}.
\end{equation}
For each bus $k$, the strategy that finds the best set $\Omega_s$
should solve the following optimization
\begin{equation}
\widehat{\Omega_s}=\arg\min\limits_{\Omega_s}{J_{\psi}({\boldsymbol \psi})},
\end{equation}
which is equivalent to minimizing the error $e_{\Omega_s}(i)$. After
the adaptation steps have been completed, the combination step is
performed as given by
\begin{equation}
\boldsymbol x_k(i)= \sum\limits_{l\in \widehat{\Omega_s}} c_{kl}
\boldsymbol\psi_l(i). \label{comb_desta}
\end{equation}
The DESTA algorithm corresponds to equations (\ref{adap
step})-(\ref{comb_desta}) and the combination weights are obtained
from (\ref{hast_rule}).

\subsection{Dynamic Sparsity-Inspired Topology Adaptation (DSITA)}

The DESTA algorithm previously described needs to examine all
possible sets to find a solution, which might result in an
unacceptable computational complexity for large systems such as the
IEEE 118--bus system \cite{Bose}. To solve this combinatorial
problem with a low complexity, we propose the sparsity--inspired
based DSITA algorithm, which bears the simplicity of a standard
diffusion LMS algorithm and is suitable for adaptive implementations
and scenarios where the parameters to be estimated are slowly
time-varying.

The zero-attracting strategy (ZA), reweighted zero–-attracting
strategy (RZA) and zero-forcing (ZF) are reported in
\cite{Chen,Meng} for sparsity aware technique. These approaches are
usually employed in applications dealing with sparse systems in such
a way that they shrink the small values in the parameter vector to
zero, which results in better convergence and steady-state
performances. Unlike existing methods that shrink the signal samples
to zero, our proposed DSITA algorithm shrinks to zero the links that
have a poor performance \cite{Xu1}.

We follow the same processing in (\ref{adap step}) for the
adaptation step, then we redesign the combination step. First, we
introduce the convex penalty term $\ell_1$--norm into the
combination step. Different penalty terms have been considered for
this task. We have adopted the heuristic approach \cite{Chen,Chen1}
called reweighted zero--attracting strategy, into the combination
step, because this strategy has shown an excellent performance and
is simple to use. Then, we consider the log-sum penalty function
\begin{equation}
{f_1(e_l(i))}= \sum\limits_{l\in \mathcal{N}_k} \log(1+\varepsilon|e_l(i)|), \label{regularization function}
\end{equation}
where the error pattern $e_l(i) (l\in \mathcal{N}_k)$ is defined as
\begin{equation}
{e_l(i)} \triangleq {z_k(i)}-{\boldsymbol h_k(i)}^*{ \boldsymbol\psi_l(i)} \label{error pattern}
\end{equation}
and $\varepsilon$ is the shrinkage magnitude. Then, the combination
step can be defined as
\begin{equation}
\boldsymbol x_k(i)= \sum\limits_{l\in \mathcal{N}_k} [c_{kl}-\rho \frac{\partial f_1(e_l(i))}{\partial e_l(i)}] \boldsymbol\psi_l(i) \label{combination step},
\end{equation}
where $\rho$ is used to control the shrinkage intensity of the
algorithm. After that, we calculate the partial derivative $e_l(i)$
of (\ref{regularization function}) by
\begin{equation}
\frac{\partial{f_1(e_l(i))}}{\partial e_l(i)}=\varepsilon\frac{{\rm{sign}}(e_l(i)) }{1+\varepsilon|\xi_{\rm min}|} \label{partial derivative}.
\end{equation}
In (\ref{partial derivative}), the parameter $\xi_{\rm min}$ stands for the minimum value
of $e_l(i)$ in each group of buses including each bus $k$ and its
neighbors. The function ${\rm{sign}}(a)$ is defined as
\begin{equation}
{{\rm{sign}}(a)}=
\left\{\begin{array}{ll}
{a/|a|}\ \ \ \ \ {a\neq 0}\\
0\ \ \ \ \ \ \ \ \ \ {a= 0}.
\end{array}
\right.
\end{equation}
To further simplify the expression in (\ref{combination step}), we
introduce the vector and matrix quantities required to describe the
combination step. We first define a vector $\boldsymbol c$ that
contains the combination coefficients for each group of buses
including bus $k$ and its neighbors as described by
\begin{equation}
{\boldsymbol c}\triangleq[c_{kl}]\ \ \ \ {l\in \mathcal{N}_k}.
\end{equation}
Then, we introduce a matrix $\boldsymbol \Psi$ that includes all the
estimated vectors, which are generated after the adaptation step in
(\ref{adap step}), for each group as given by
\begin{equation}
{\boldsymbol \Psi}\triangleq[\boldsymbol \psi_l(i)]\ \ \ \ {l\in \mathcal{N}_k}.
\end{equation}
An error vector $\boldsymbol e$ that contains all the error values
calculated through (\ref{error pattern}) for each group is expressed
by
\begin{equation}
{\boldsymbol e}\triangleq[e_l(i)]\ \ \ \ {l\in \mathcal{N}_k}. \label{error vector}
\end{equation}
To devise the sparsity--inspired approach, we have modified the
vector $\boldsymbol e$ in the following way: the maximum value
$e_l(i)$ in $\boldsymbol e$ will be set to $|e_l(i)|$; the minimum
value $e_l(i)$ will be set to $-|e_l(i)|$, while the remaining
entries will be set to zero. Finally, by inserting (\ref{partial
derivative})--(\ref{error vector}) into (\ref{combination step}),
the combination step will be changed to
\begin{equation}
\begin{split}
\boldsymbol x_k(i)&={\sum\limits_{j=1}^{\mathcal{N}_k} [\boldsymbol c_j-\rho \frac {\partial f_1}{\partial \boldsymbol e_j}(\boldsymbol e_j)] \boldsymbol\Psi_j}\\
&={\sum\limits_{j=1}^{\mathcal{N}_k} [\boldsymbol c_j-\rho
\varepsilon\frac{{\rm{sign}}({\boldsymbol e_j})
}{1+\varepsilon|\xi_{\rm min}|}] \boldsymbol\Psi_j}  \label{combination step2}.
\end{split}
\end{equation}
The proposed DSITA algorithm performs dynamic topology adaptation by
the adjustment of the combination coefficients through $\boldsymbol
c$ in (\ref{combination step2}). For the neighbor bus with the
largest MSE value, after our modifications for $\boldsymbol e$, its
$e_l(i)$ value in $\boldsymbol e$ will be a positive number which
will lead to the term $\rho\varepsilon\frac{{\rm{sign}}({\boldsymbol
e_j})}{1+\varepsilon|\xi_{\rm min}|}$ in (\ref{combination step2})
being positive too. This means that the combining coefficient for
this bus will be reduced and the weight for this bus to build the
$\boldsymbol x_k(i)$ is reduced too. In contrast, for the neighbor
bus with the minimum MSE, as its $e_l(i)$ value in $\boldsymbol e$
will be a negative number, the term
$\rho\varepsilon\frac{{\rm{sign}}({\boldsymbol
e_j})}{1+\varepsilon|\xi_{\rm min}|}$ in (\ref{combination step2})
will be negative too. As a result, the weight for this node
associated with the minimum MSE to build the $\boldsymbol x_k(i)$ is
increased. For the remaining neighbor buses, the $e_l(i)$ value in
$\boldsymbol e$ is zero, which means the term
$\rho\varepsilon\frac{{\rm{sign}}({\boldsymbol
e_j})}{1+\varepsilon|\xi_{\rm min}|}$ in (\ref{combination step2})
is zero and there is no change for their weights to build the
$\boldsymbol x_k(i)$. The constraint on the combination of the
coefficients in (\ref{ctotal 1}) is still satisfied. In conclusion,
each bus $k$ will first obtain an local estimate through (\ref{adap
step}). Then, each bus will employ (\ref{error pattern})-
(\ref{combination step2}) to perform the dynamic topology
adaptation.

\section{Simulations}

\begin{figure}[!htb]
\begin{center}
\def\epsfsize#1#2{0.625\columnwidth}
\epsfbox{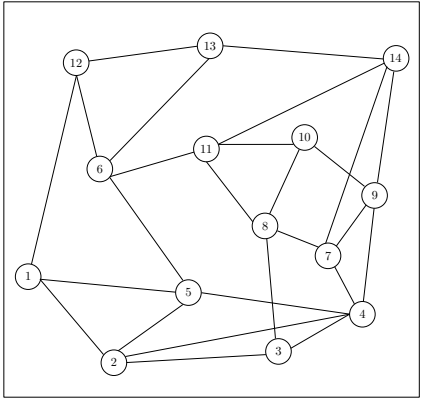} \vspace{-0.5em} \caption{\footnotesize
IEEE 14--bus system for simulation}
\vskip -5pt
\label{fig2}
\end{center}
\end{figure}

In this section, we compare our proposed dynamic topology adaptation
algorithms, DESTA and DSITA, with the $\mathcal{M}$--$\mathcal{CSE}$
\cite{Xie} and traditional diffusion ATC algorithm \cite{Lopes2}
based on the MSE performance and the Phase Angle Gap. The MSE
comparison is used to determine the accuracy of the algorithms, and
the Phase Angle Gap is used to compare the convergence rate. In our
scenario, 'Phase Angle Gap' stands for the phase angle difference
between the target $\boldsymbol \theta$ and the estimate
$\boldsymbol x$ for all buses. We define the IEEE--14 bus system as
in Fig. \ref{fig2}.

\begin{figure}[!htb]
\begin{center}
\def\epsfsize#1#2{1\columnwidth}
\epsfbox{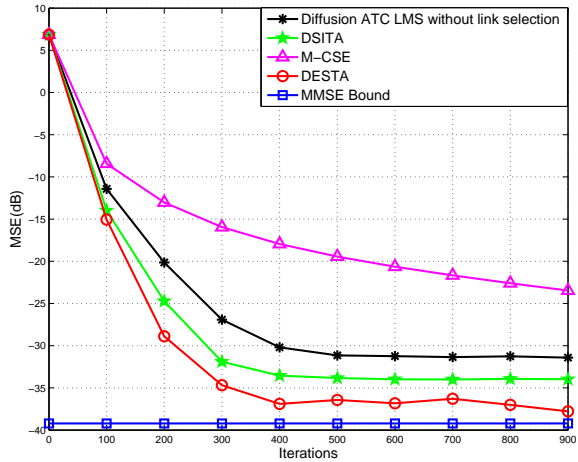} \vspace{-1.5em} \caption{\footnotesize
MSE  performance curves.} \vskip -5pt \label{fig3}
\end{center}
\end{figure}

\begin{figure}[!htb]
\begin{center}
\def\epsfsize#1#2{1\columnwidth}
\epsfbox{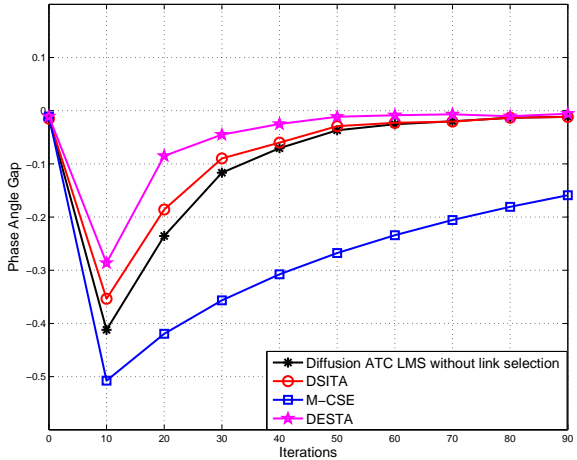} \vspace{-1.5em}
\caption{\footnotesize
Phase Angle Gap Comparison For Node 5.} \vskip -5pt \label{fig4}
\end{center}
\end{figure}

All buses are corrupted by additive white Gaussian noise with equal
variance $\sigma^2=0.001$. The step size for the proposed DESTA and
DSITA algorithms is 0.018. The parameter vector $\boldsymbol \theta$
is set to an all-one vector. The sparsity parameters of the DSITA
algorithm are set to $\rho=0.07$ and $\varepsilon=10$. The results
are averaged over 100 independent runs. From Fig. \ref{fig3}, it can
be seen that our proposed DESTA algorithm has the best performance,
and significantly outperforms the standard diffusion ATC algorithm
and $\mathcal{M}$--$\mathcal{CSE}$ algorithm. DSITA is slightly
worse than DESTA, which outperforms the remaining techniques.

To compare the convergence rate, we use the term -- 'Phase Angle
Gap' to describe the results. We pick bus 5 and the first 90
iterations as an example to show our results. In Fig. \ref{fig4},
the DESTA algorithm still has the fastest convergence rate, while
the DSITA algorithm is the second fastest. The estimates
$\boldsymbol x$ made from our proposed dynamic topology adaptation
algorithms can quickly reach the target $\boldsymbol \theta$, which
means the Phase Angle Gap will converge to zero.

\section{Conclusion}

In this paper, two dynamic topology adaptation strategies have been
proposed for distributed estimation in smart grids. The DESTA
algorithm uses an exhaustive search to perform the dynamic topology
adaptation, and DSITA employs a sparsity--inspired approach with the
$\ell_1$--norm penalization. Numerical results have shown that the
two proposed algorithms achieve a better convergence rate and lower
MSE values than the existing distributed state estimation
algorithms. These results hold also when employing other algorithms
including RLS and distributed CG \cite{Xu} techniques. 

{\footnotesize
\bibliographystyle{IEEEbib}
\bibliography{reference}}

\end{document}